\documentclass[sigconf, screen=true, review=false, anonymous=false, authordraft=False, language=english]{acmart}
\usepackage{blindtext}
\usepackage{subfig}

\AtBeginDocument{%
  }

\setcopyright{cc}

\setcctype{by}

\acmConference[MuC'26]{Mensch und Computer 2026 – Tagungsband, Gesellschaft für Informatik e.V.}{30. August - 02. September 2026}{Duisburg, Germany}
\acmDOI{10.18420/muc2026-mci-ws103-310}

\begin{document}

\title[A Formative Usability Study of a Conversational Product Advisor]{Transparent by Design, Usable in Practice? A Formative Usability Study of a Conversational Product Advisor}


\settopmatter{authorsperrow=3}


\author{Kevin Schott}
\orcid{0009-0005-7858-9462}
\affiliation{%
 \institution{GESIS – Leibniz Institute for the Social Sciences}
 \city{Cologne}
 \country{Germany}
 }
\email{kevin.schott@gesis.org}

\author{Daniel Hienert}
\orcid{0000-0002-2388-4609}
\affiliation{%
 \institution{GESIS – Leibniz Institute for the Social Sciences}
 \city{Cologne}
 \country{Germany}
 }
\email{daniel.hienert@gesis.org}

\author{Dagmar Kern}
\orcid{0000-0003-1794-625X}
\affiliation{%
 \institution{GESIS – Leibniz Institute for the Social Sciences}
 \city{Cologne}
 \country{Germany}
 }
\email{dagmar.kern@gesis.org}

\begin{abstract}
Large language models can make conversational product advisors fluent but opaque. If they hide the logic behind a ranking and the evidence for a recommendation inside natural-language replies, they challenge users' ability to understand, trust, and steer the results. One response is to build transparency into the advisor. We report a formative, moderated think-aloud usability study of one such system: a chatbot for laptop search with constrained natural-language generation, an on-demand ranking explanation, and a comparison feature. Seven participants completed three laptop-search tasks and reported post-task usability measures. We coded their sessions into severity-rated usability problems. Ease and satisfaction during the tasks were high, but two findings stand out. First, transparency by design did not guarantee understanding: several participants valued the ranking explanation in principle, yet it caused the most severe problem. Second, participants valued the effort the advisor saved, but some wanted additional direct-manipulation controls. We contribute a severity-prioritized set of usability problems and design implications for human-centered conversational product advisors.
\end{abstract}

\ccsdesc[500]{Information systems~Users and interactive retrieval}
\ccsdesc[500]{Human-centered computing~Usability testing}

\keywords{Human-Centered AI, conversational recommender systems, usability testing, explainability}

\maketitle

\section{Introduction and Background}
Conversational recommender systems support users through a multi-turn dialogue in which the system elicits preferences, suggests items, and explains its suggestions~\cite{jannach2021survey, gao2021advances}, letting users iteratively revise their requirements, for example by critiquing the items shown~\cite{chen2012critiquing}. Radlinski and Craswell~\cite{radlinski2017framework} frame this as revealment: user revealment helps users express or discover their needs, while system revealment exposes what the system can and cannot do. This flexibility has a cost, however: when results are delivered as fluent text, the ranking logic and the sources of the information are easily concealed, even though users need them to judge reliability and relevance~\cite{lajewska2024explainability}. Keeping the interaction natural while giving users enough insight to trust and act on the results is thus an open design problem.

In a product advisor, the system recommends while the person remains the decision-maker. Human-Centered AI holds that supporting this role need not be a trade-off: rather than treating more automation as necessarily less human control~\cite{sheridan2011automation}, it seeks to combine high automation with high human control to make automation reliable, safe, and trustworthy~\cite{shneiderman2022hcai, shneiderman2020rst}. Interaction design has long contrasted direct manipulation, where users keep visible control, with delegation to software agents~\cite{shneiderman1997dm}. Mixed-initiative approaches sought to couple the two~\cite{horvitz1999mixed}. A conversational advisor sits on the delegation side, and the extent of control it should afford also depends on context, users' cognitive limits, and individual differences~\cite{pacailler2022hcai}. Explanations are a means to understanding but sometimes serve conflicting aims, and greater transparency can raise or lower trust depending on the confidence it gives~\cite{holzinger2019causability, dwivedi2023xai, tintarev2012explanations}. Transparency in practice also reaches beyond technical explainability: even a fully explainable system can leave people unable to make sense of, or act on, a decision~\cite{mckay2022transparency}. Ultimately, whether users rely on a system depends on their trust in it~\cite{koerber2018trust, johnson2026trust}, and a hazard specific to LLM-based assistants is sycophancy, in which human-feedback-tuned models align with a user's stated views over the truth~\cite{sharma2024sycophancy}.

The system we study~\cite{cleo2026} takes this route, building transparency into the advisor itself: an auditable, per-attribute ranking that it can explain on demand, generation constrained to a product catalog, and decision-support features for inspecting and comparing options. Yet designing for transparency does not guarantee that it works in practice. Accordingly, we raise two questions: (1) Is the built-in transparency actually perceived, understood, and trusted by users? (2) If users partially delegate the search to a chatbot, do they still have the level of control they expect? We address these questions through a formative think-aloud study with seven participants, yielding severity-rated usability problems and concrete design implications for human-centered conversational product advisors.

The remainder of this paper introduces the evaluated system (Section~\ref{sec:system}), details our method (Section~\ref{sec:method}), reports the results (Section~\ref{sec:results}), and discusses the findings, design implications, limitations, and future work (Sections~\ref{sec:discussion} and~\ref{sec:conclusion}).

\section{Evaluated System}\label{sec:system}

\begin{figure*}[!t]
  \centering
  \subfloat[Elicitation: opening prompt and the user's request (top of chat)]{%
    \includegraphics[width=0.495\textwidth]{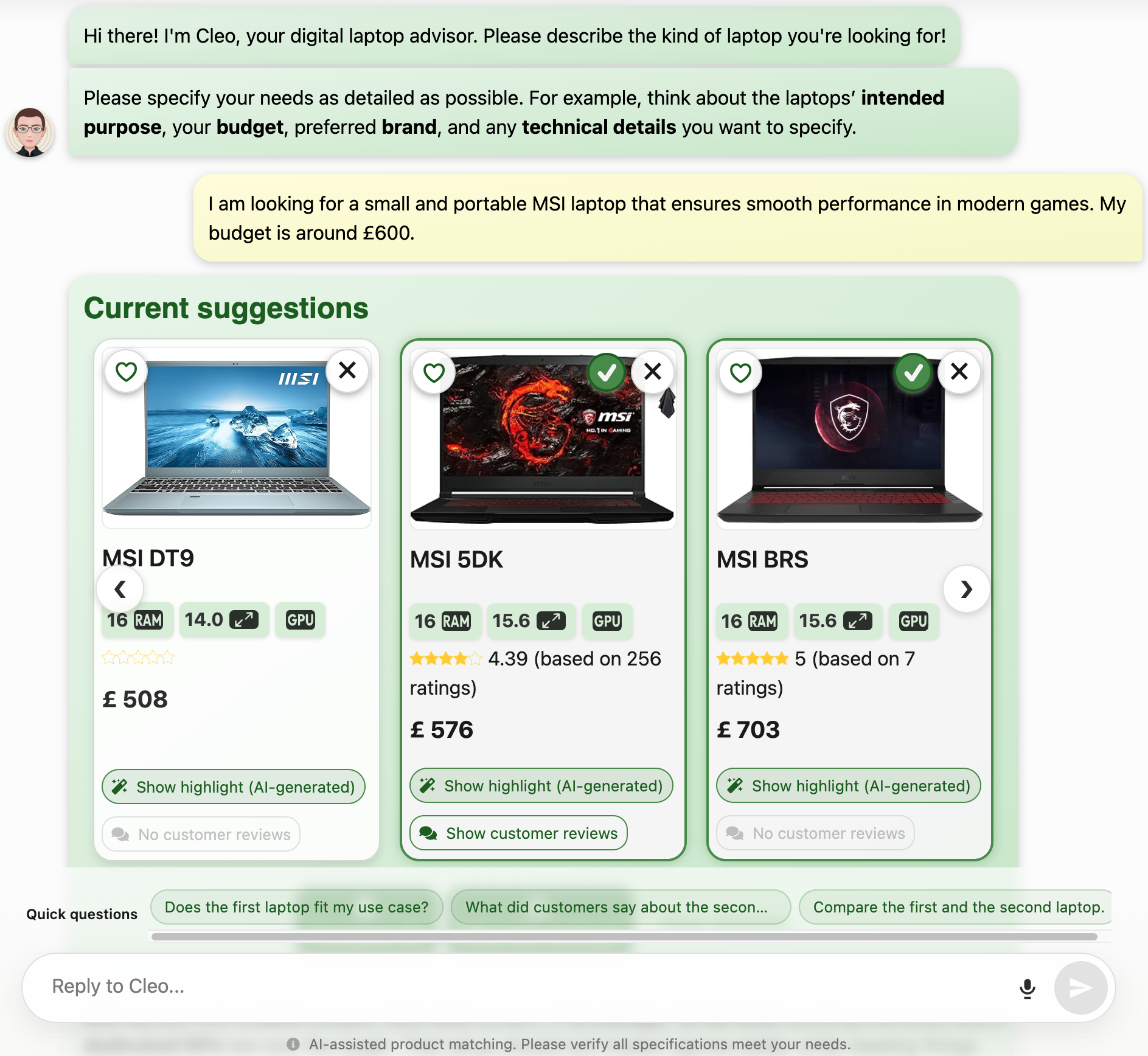}%
    \label{fig:chat_up}%
    \Description{The top of the chat. Cleo, the advisor, greets the user and asks them to describe the laptop they want, prompting for intended purpose, budget, preferred brand, and technical details. The user asks for a small, portable MSI laptop for smooth performance in modern games with a budget of around 600 pounds. A Current suggestions panel begins to appear below the conversation.}%
  }
  \hspace{0.1cm}
  \subfloat[Carousel and the advisor reflecting back the inferred requirements (scrolled down)]{%
    \includegraphics[width=0.495\textwidth]{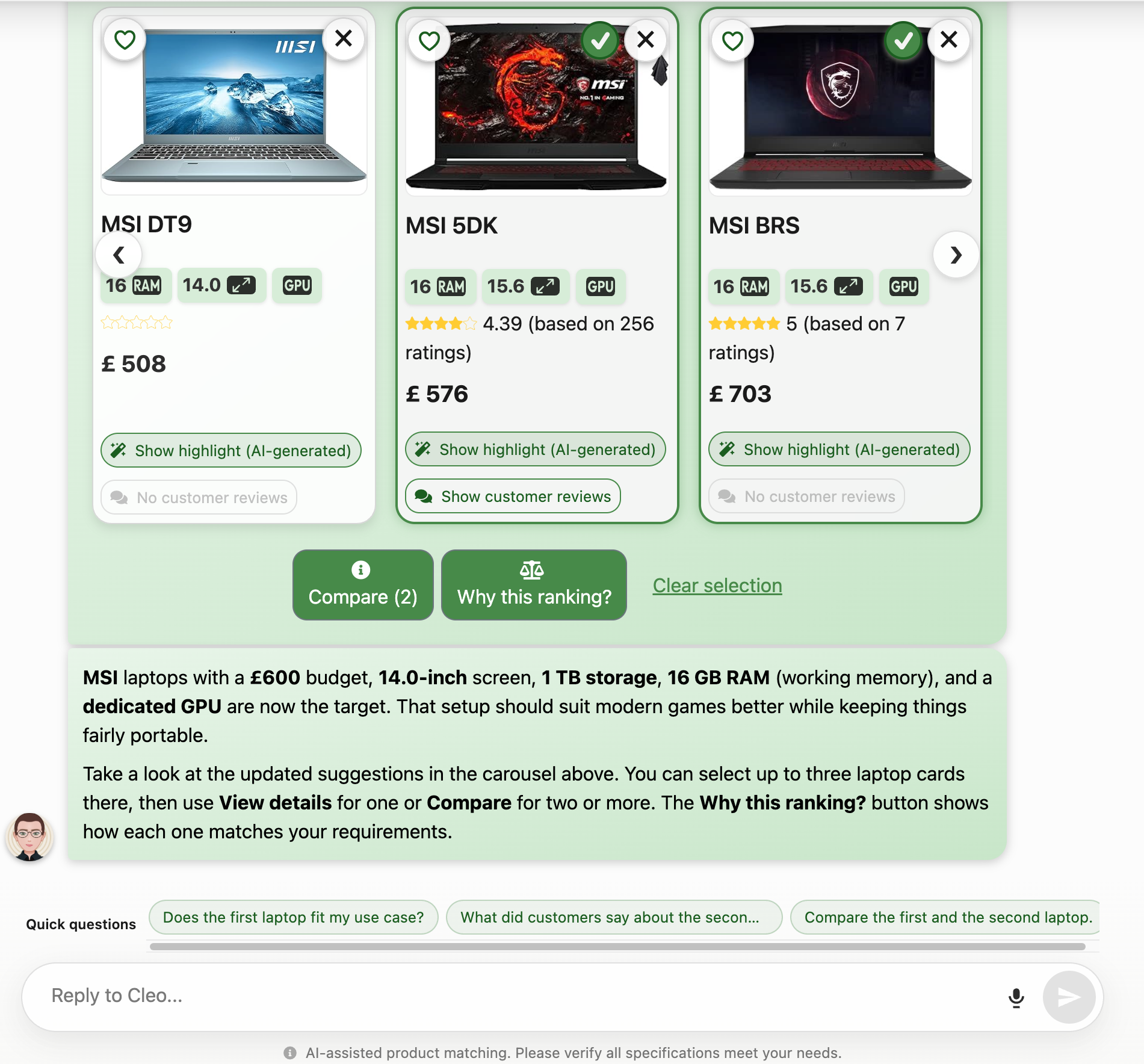}%
    \label{fig:chat_down}%
    \Description{The same chat scrolled down. A carousel shows three MSI laptop cards (MSI DT9, MSI 5DK, MSI BRS) with RAM, screen-size and GPU tags, star ratings, and prices; two cards are selected with green check marks, above Compare and Why this ranking? buttons. Below, the advisor reflects back the inferred requirements: an around-600-pound budget, a 14-inch screen, 1 TB storage, 16 GB RAM, and a dedicated GPU, and points to the View details, Compare, and Why this ranking? actions.}%
  }
  \\[0.5ex]
  \subfloat[``Why this ranking?'': radar match view and written summaries]{%
    \raisebox{-0.5\totalheight}{\includegraphics[width=0.495\textwidth]{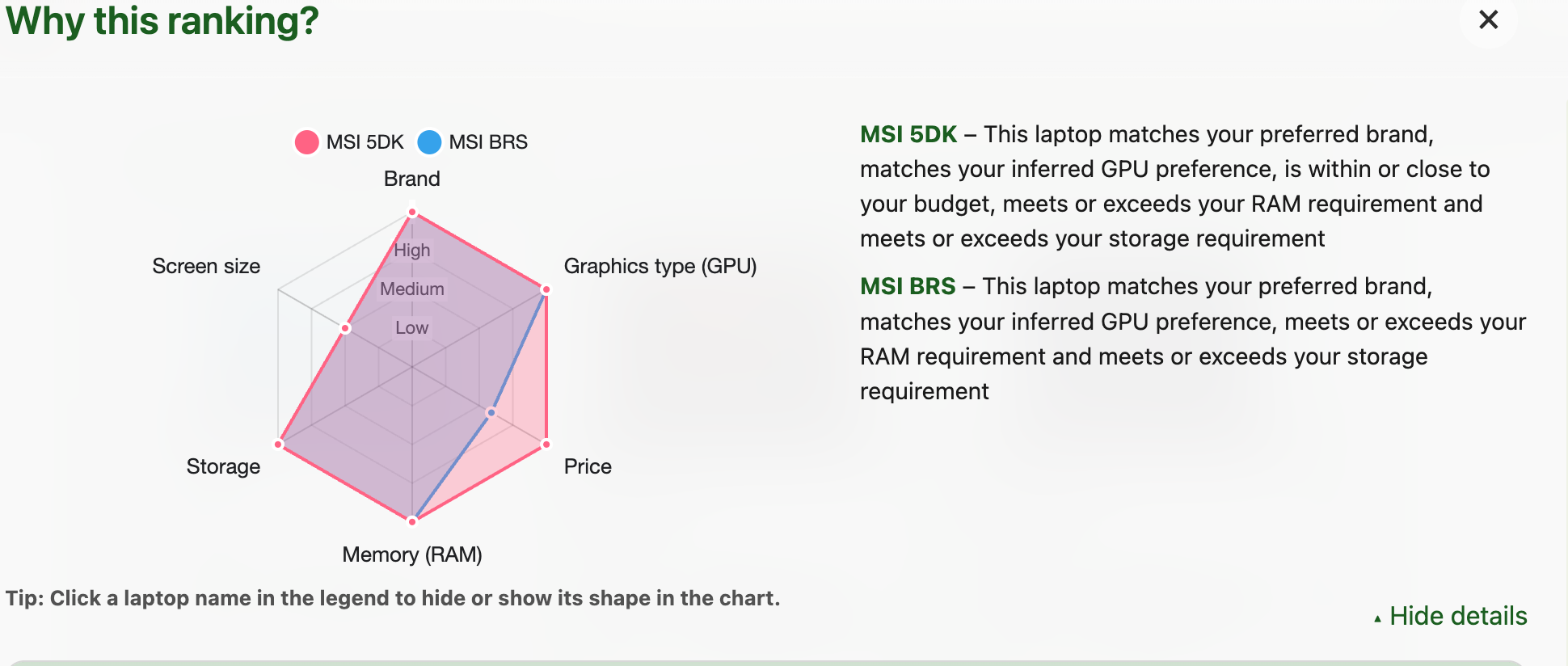}}%
    \label{fig:rank_up}%
    \Description{The upper part of the Why this ranking? modal. A radar chart compares the MSI 5DK and MSI BRS across brand, graphics type, price, memory, storage, and screen size, next to a written summary stating that each laptop matches the preferred brand and inferred GPU preference and meets or exceeds the RAM and storage requirements.}%
  }
  \hspace{0.1cm}
  \subfloat[``Why this ranking?'': per-attribute penalty table and legend]{%
    \raisebox{-0.5\totalheight}{\includegraphics[width=0.495\textwidth]{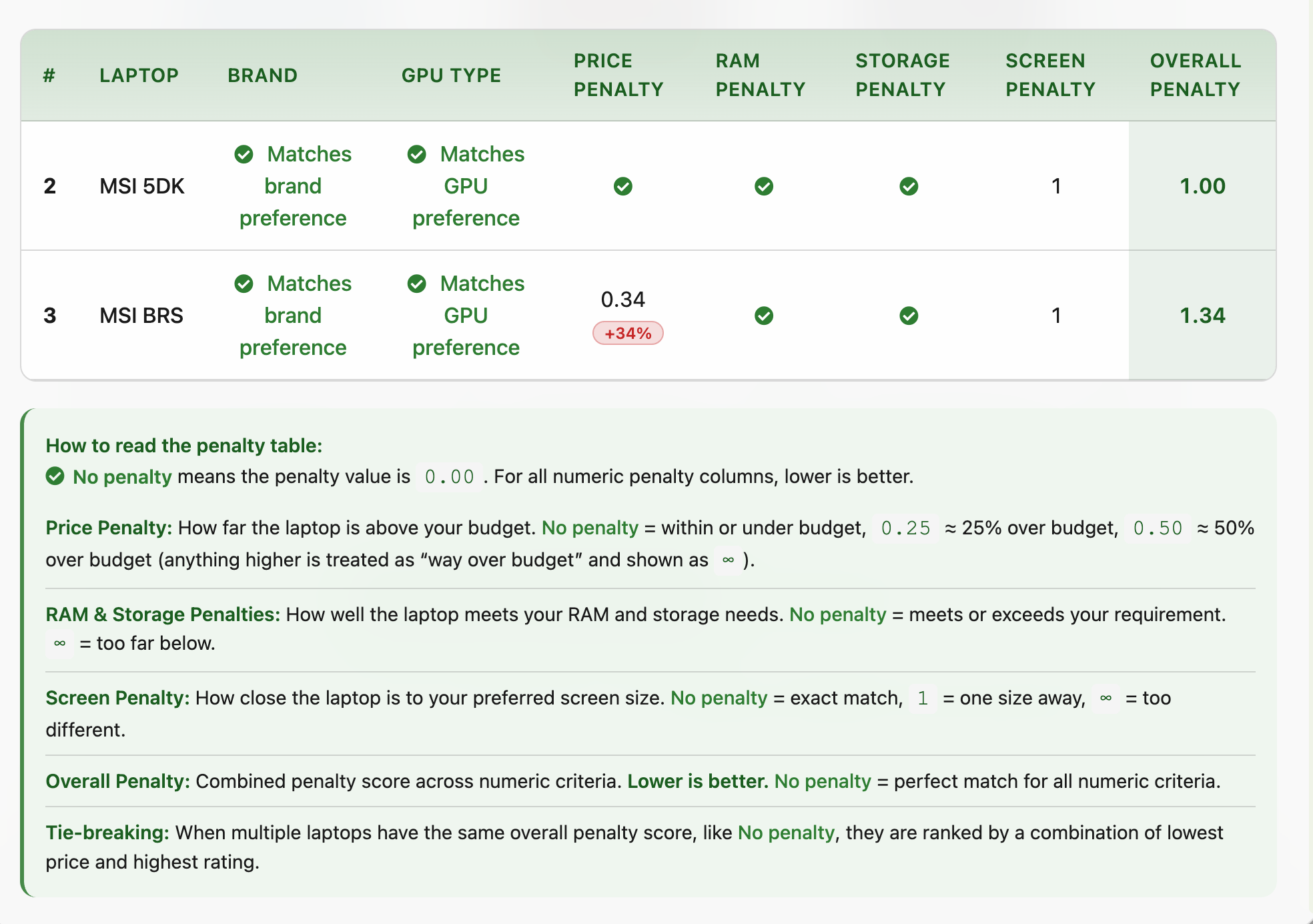}}%
    \label{fig:rank_down}%
    \Description{The lower part of the Why this ranking? modal. A table lists per-attribute penalties and an overall penalty, with the MSI 5DK at 1.00 and the MSI BRS at 1.34, including a price penalty of 0.34 marked as plus 34 percent over budget. A legend explains that lower penalties are better and defines the price, RAM, storage, screen, and overall penalties and the tie-breaking rule.}%
  }
  \caption{The interface of the evaluated product advisor and the views central to our findings: (a, b) the conversational elicitation and the advisor reflecting back the inferred requirements alongside the recommendation carousel; (c, d) the on-demand ranking explanation, split into the radar-based match view and the per-attribute penalty table with its legend. (Continued on next page)}
  \label{fig:ui}
\end{figure*}

\begin{figure*}[!t]
  \ContinuedFloat
  \centering
  \subfloat[Comparison modal (``View Details'' / ``Compare'')]{%
    \includegraphics[width=0.90\textwidth]{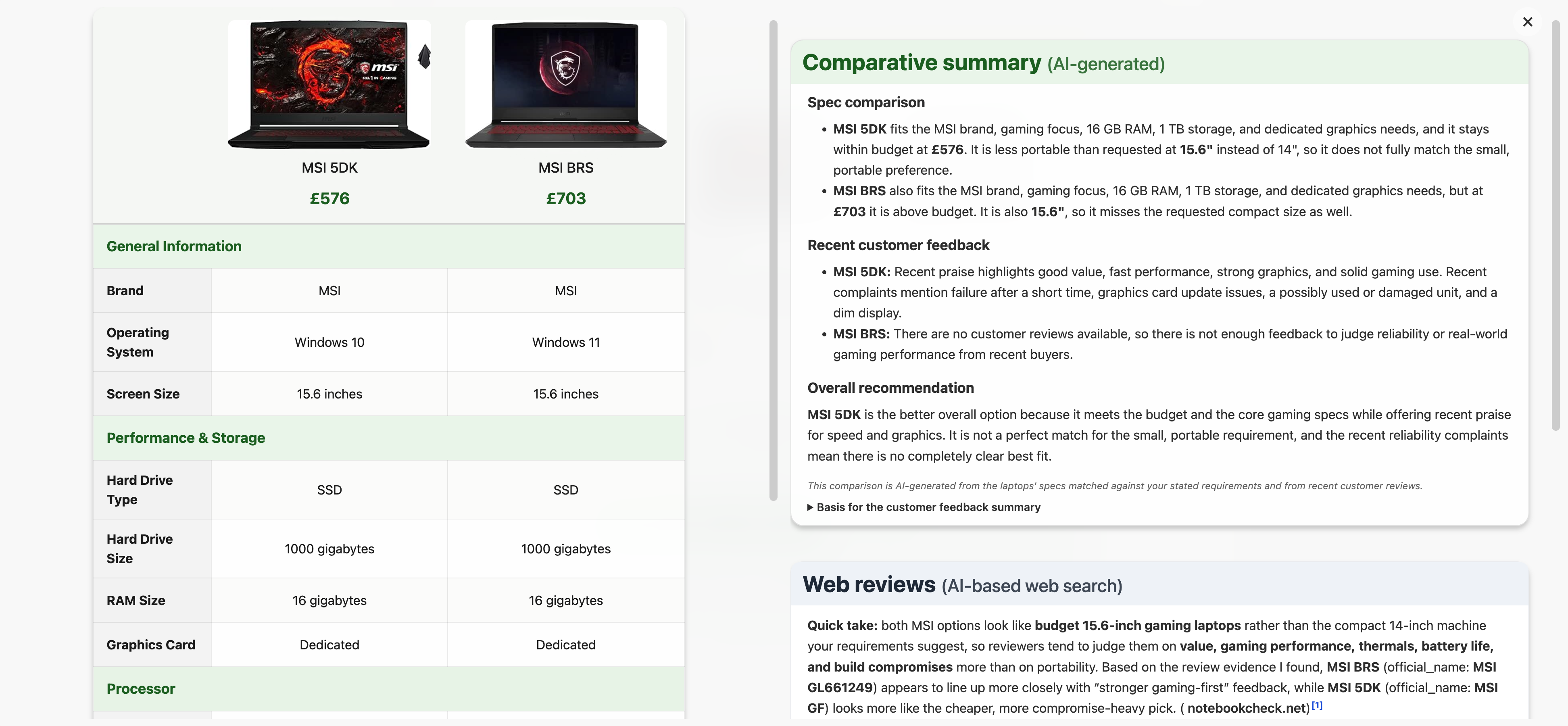}%
    \label{fig:compare}%
    \Description{The comparison modal. On the left, a specification table compares the MSI 5DK and MSI BRS on brand, operating system, screen size, drive type and size, RAM, and graphics card. On the right, an AI-generated comparative summary, a recent-customer-feedback summary, and an overall recommendation favoring the MSI 5DK, followed by an AI-based web-review search. A note states that the summary is AI-generated from the specifications and recent reviews.}%
  }
  \caption{The interface of the evaluated product advisor (continued): (e) the comparison modal.}
\end{figure*}

The system under study~\cite{cleo2026} is a conversational product advisor (chatbot) for laptops, designed for transparency. Figure~\ref{fig:ui} shows the interface and the views central to our findings.\footnote{The full task descriptions, interview guide, questionnaire items, and interface screenshots are provided as supplementary material on OSF: \url{https://doi.org/10.17605/OSF.IO/S3K4B}.} It uses a hybrid architecture that separates a deterministic ranker from a constrained language model: the ranker applies categorical filters and numeric loss functions to a catalog of product specifications, while the language model generates descriptions grounded in that catalog, limiting hallucinated or persuasive content. Users interact in natural language via text or speech, and the advisor maps their explicitly or implicitly stated requirements (for example, a mentioned purpose of use) to the filter attributes brand, price, screen size, storage, RAM, and integrated versus dedicated GPU. Recommendations appear in a carousel of nine laptops, beneath which a ``Why this ranking?'' button opens an explanation modal for up to three selected laptops. The modal presents a radar chart on the left showing how well each option matches these attributes, next to a written account of the matches on the right. Clicking ``View details'' inside the modal extends it with a table of the categorical filter matches and the exact loss (penalty) values the ranker computes for each attribute, with a legend explaining the penalties. A second button beneath the carousel adapts to the selection, reading ``View details'' for one laptop and ``Compare (n)'' for two or three, and opens a modal with a specification table, a summary of specifications and recent customer reviews, and a web-review search. For transparency, the advisor also reflects back how it interpreted a message that introduces or changes preferences, for example translating a wish to play modern games into a dedicated-GPU requirement. It further labels all AI-generated content with its provenance, flagging the results of the web-review search as possibly incomplete or inaccurate. For the study, it was connected to a catalog of Windows laptops derived from Amazon data.

\section{Method}\label{sec:method}
To answer both questions, we combined observed usage with self-reported measures in a task-based usability test.

\subsection{Study Design and Procedure}
We conducted a formative, moderated, remote think-aloud usability test to identify problems and requirements for future iterations of the advisor. Participants verbalized their thoughts, expectations, and confusions in individual Microsoft Teams sessions that were screen- and audio-recorded and automatically transcribed. Each session lasted approximately 40 minutes and followed a fixed structure. After providing consent, participants completed the three main tasks (described below), followed by a post-test questionnaire hosted on SoSci Survey\footnote{\url{https://www.soscisurvey.de/en/index}} and a semi-structured interview. They then answered demographic questions on age and gender, and two further seven-point items: their self-rated knowledge of laptops (``How would you classify your knowledge of laptops?'', with the instruction to consider aspects such as storage, RAM, and GPU, from 1 = no knowledge or non-expert, to 7 = high knowledge or expert) and their perceived reliability of AI-generated content (``How reliable do you find AI-generated content in general?'', from 1 = not at all reliable, to 7 = very reliable). The session closed with a short debriefing. A task ended when the participant identified a suitable option, decided to stop searching, or reached the roughly seven-minute time limit.

\begin{table}[t]
  \centering
  \small
  \caption{The three tasks and the respective chatbot capability under test.}
  \Description{A table listing the three study tasks with their input type and the capability each one evaluates.}
  \label{tab:tasks}
  \begin{tabular}{@{}l p{0.30\linewidth} p{0.35\linewidth}@{}}
    \toprule
    Task & Input & Capability under test \\
    \midrule
    1. Free search
      & Open, the participant's own requirements
      & Natural-language preference elicitation and initial UI exploration \\[2pt]
    2. Gaming search
      & Fixed prompt (MSI gaming laptop, approx.\ \pounds{}600)
      & The ``Why this ranking?'' ranking-explanation feature \\[2pt]
    3. Work search
      & Fixed prompt (Lenovo 16-inch work laptop, \pounds{}800), then a
        refinement (HP, fast, under \pounds{}600, more storage)
      & Context retention across refinement and the ``Compare'' feature \\
    \bottomrule
  \end{tabular}
\end{table}

Participants worked through three tasks in a fixed order (full prompts in the OSF materials). Task~1 was an open warm-up with the participant's own requirements, exercising natural-language elicitation and initial UI exploration. Task~2 used a fixed prompt for an MSI gaming laptop costing about \pounds{}600, and participants selected the top match plus two lower-ranked options and explored the ``Why this ranking?'' explanation. Task~3 began from a fixed prompt for a 16-inch Lenovo work laptop costing \pounds{}800. Mid-task, the requirements changed (a fast HP device under \pounds{}600 with more storage), after which participants selected three laptops, used the ``Compare'' feature, and chose one for their wishlist.

\subsection{Measures}
After each task, participants answered two single-item measures on seven-point Likert scales: the Single Ease Question (SEQ, \textit{``Overall, how difficult or easy did you find this task?''})~\cite{sauro2009seq} and a question on satisfaction with the interaction (\textit{``How satisfied were you with the chatbot interaction during the task?''}). After completing all tasks, they filled in the shortened BUS-11 ChatBot Usability Scale~\cite{borsci2023bus, borsci2024bus11}. We omitted the \textit{Accessibility} and \textit{Privacy} items as not applicable and kept only those of its \textit{Functional and Interactive Conversation Quality} and \textit{Responsiveness} factors. A closing semi-structured interview asked about the most helpful and most confusing parts of the interaction, the helpfulness of the ranking explanation and comparison feature when used, and whether and why participants would trust or use the chatbot for a real search. All user and assistant messages were logged, and the moderator took notes.

\subsection{Analysis}
Our analysis was primarily qualitative. We drew on the screen recordings, automatic transcripts, message logs, and moderator notes to code usability problems and additional emerging themes. Coding followed a thematic analysis~\cite{braun2006thematic} that combined inductive coding with a priori candidate themes, including requirement interpretation, transparency of ranking, discoverability, and trust. Each problem received a severity rating from 1 (minor) to 4 (critical). Following Nielsen~\cite{nielsen1994heuristic}, who defines severity by frequency, impact, and persistence, we based the ratings on the number of participants affected, the impact (on task success, effort, or trust), and whether the problem could be overcome without help. The first author did the coding and rating. The quantitative measures are reported descriptively.

\subsection{Participants}
Because we aimed to surface key usability problems rather than precisely estimate their frequency, we used a small sample. Nielsen and Landauer~\cite{nielsen1993model} show strongly diminishing returns in usability problem detection and put the optimal number of test users for a small-scale project at seven, the size we adopted. We recruited seven participants through Prolific\footnote{\url{https://www.prolific.com/}}, screening for UK residence and English fluency, and compensated each with \pounds{}10, about \pounds{}15 per hour. A form-factor check confirmed that every participant used a laptop or desktop. The sample comprised four women and three men, aged 24 to 71 ($M=40.7,\, SD=16.6,\, Mdn=35$). On seven-point scales, self-rated knowledge of laptops was $M=4.3,\, SD=1.3$, and perceived reliability of AI-generated content $M=5.6,\, SD=1.3$.

\section{Results}\label{sec:results}
Overall, participants rated the advisor favorably, yet the think-aloud data surfaced seven usability problems, with the ranking explanation as the most severe.

\subsection{Quantitative Results}

\begin{figure*}[t]
  \centering
  \includegraphics[width=0.9\textwidth]{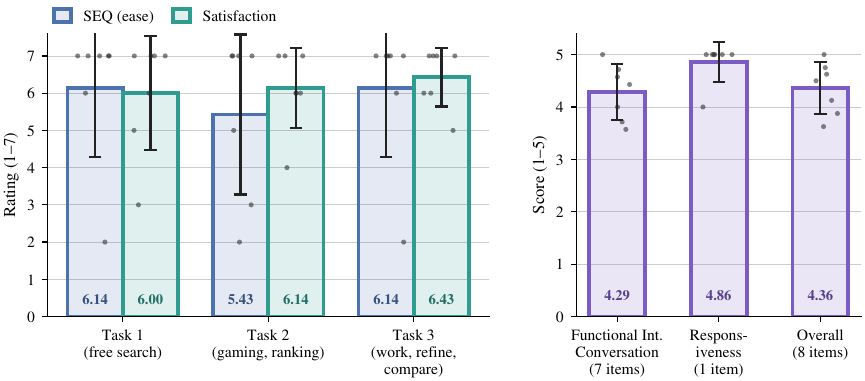}
  \caption{Quantitative results ($N=7$; bars = mean, whiskers = $SD$, dots = individual participants). Left: post-task ease (SEQ) and satisfaction per task on seven-point scales. Right: BUS-11 scores on five-point scales for the administered \textit{Functional Interactive Conversation} and \textit{Responsiveness} items.}
  \Description{A two-panel bar chart. Left panel: SEQ and satisfaction ratings for the three tasks on a one-to-seven scale. SEQ means are 6.14, 5.43, and 6.14; satisfaction means are 6.00, 6.14, and 6.43. Task 2 has the lowest and most variable ease ratings. Right panel: BUS-11 scores on a one-to-five scale, with Functional Interactive Conversation at 4.29, Responsiveness at 4.86, and Overall at 4.36. Standard-deviation whiskers and individual participant dots are overlaid on all bars.}
  \label{fig:quant}
\end{figure*}

Across the three tasks, self-reported ease and satisfaction were consistently high (Figure~\ref{fig:quant}). Task~2, centered on the ranking explanation, had the lowest and most variable ease ratings, while satisfaction rose slightly across the tasks (Figure~\ref{fig:quant}, left). Across the administered BUS-11 items (rated on five-point scales), the mean was $M=4.36,\, SD=0.50$, with the \textit{Responsiveness} factor highest ($M=4.86,\, SD=0.38$) and \textit{Functional Interactive Conversation} at $M=4.29,\, SD=0.53$. Descriptively, the two lowest-scoring items of this factor concerned whether the chatbot gave only the information needed ($M=3.57$) and kept track of context ($M=4.14$), converging with the desire for more tailoring and the context-retention problem below.

\subsection{Usability Problems}

\begin{table*}[t]
  \centering
  \caption{Usability problems by severity (1 = minor, 4 = critical), then by participants raising the issue ($n$ of 7).}
  \Description{A table of seven usability problems with their severity rating, the number of participants who raised each, and a brief description of the supporting evidence.}
  \label{tab:problems}
  \begin{tabular}{@{}p{0.29\linewidth}cc p{0.53\linewidth}@{}}
    \toprule
    Problem & Sev. & $n$ & Evidence \\
    \midrule
    Ranking explanation hard to interpret & 3 & 4 & Opaque, negatively framed penalty values and a radar chart hard to read with three laptops \\
    Discoverability and navigation & 2 & 5 & Unclear card selection and Compare vs.\ Why-this-ranking confusion \\
    Desire for more control and tailoring & 2 & 4 & Wanted explicit filters and user-directed sorting, and responses tailored to the request \\
    Ranking order not signaled & 2 & 2 & Unclear that the leftmost item is the best ranked \\
    Web-review credibility & 2 & 2 & AI web reviews felt bland and unattributed \\
    Voice input cuts off after a pause & 2 & 2 & Microphone stopped capturing mid-utterance for voice users \\
    Limited context retention & 2 & 2 & Lost the referent of just-discussed models \\
    \bottomrule
  \end{tabular}
\end{table*}

Thematic analysis identified seven usability problems, summarized in Table~\ref{tab:problems} and ordered by severity and then by prevalence. The only serious problem concerned the ``Why this ranking?'' explanation. Several participants valued it in principle. \emph{P7} called the ranking \textit{``clear,''} and \emph{P3} said \textit{``the ranking system in particular I thought was really good,''} but its presentation blocked comprehension for others. Because the per-attribute values are framed as penalties, they read as negative: \emph{P5} observed that \textit{``the penalty is a negative thing''} whereas \textit{``I want a positive experience,''} and that a higher number \textit{``points me in the wrong direction.''} \emph{P4} could not tell what the values referred to, asking \textit{``why is the price penalty given as 34\%?''}, and would have preferred raw numbers to the system \textit{``inventing penalties.''} The radar chart was hard to read with three laptops overlaid, and \emph{P4} felt \textit{``a bar chart would be easier.''} The penalties covered inferred requirements, not only explicitly stated ones, confusing \emph{P4} when an unrequested storage criterion lowered a laptop's rank.

The most widespread problem was discoverability and navigation, raised by five of the seven participants. The purpose of selecting a card was unclear. As \emph{P4} put it, \textit{``I can select, I can unselect, but it doesn't really do me anything.''} Three participants opened the comparison modal when they had meant to open the ``Why this ranking?'' explanation and needed redirecting. When a new message appeared below the fold after a participant had scrolled up in the chat history, nothing signaled it, so \emph{P5} briefly lost the response, asking \textit{``Where did all that typing go?''}

The remaining five problems were also moderate. Some participants wanted more control over and tailoring of the interaction. \emph{P1} asked for explicit filter controls, \textit{``some kind of simple filter that you can find it everywhere like for example on Amazon,''} while \emph{P5} wanted to direct the sorting, for instance to \textit{``rank these in price order, descending,''} or to sort by storage or RAM. Others wanted responses adapted to a request or use case, such as a simpler explanation or a gaming-specific graphics rating (\emph{P3}, \emph{P6}). Two participants were unsure how to read the ranking order, with \emph{P4} \textit{``not completely certain how I'm meant to interpret this,''} wondering whether the cards were to be read \textit{``left to right''} as best first. To \emph{P4}, who wanted professional rather than crowd-sourced opinions, the AI web-review summaries felt \textit{``bland''} and unattributed, although \emph{P3} valued that sources were cited, finding it \textit{``always good to see some sources being put in there, especially in terms of user trust.''} For the two participants who dictated their input, the microphone stopped capturing after a short pause and its button had to be re-pressed (\emph{P2}, \emph{P4}). Finally, context retention was limited: the system occasionally lost track of items it had just discussed, for instance failing to resolve \emph{P6}'s request for \textit{``the reviews for these particular models''} one turn after listing them.

\subsection{Positive Experience and Trust}
Participants were positive in several respects. Responses were perceived as fast (five of seven), and the card-based visual presentation was well received, with \emph{P2} noting it \textit{``wasn't just giving you text.''} The comparison feature was valued for reaching a decision. Six of seven would use or try such an advisor, with four citing saving effort or time versus a conventional search: \emph{P2} would \textit{``prefer to use a chatbot than to do any search in comparison myself,''} as it was \textit{``saving me a lot of time.''} On trust, transparency and a neutral tone were reassuring. \emph{P3} found the answers \textit{``transparent and fair''} with no apparent \textit{``agenda or narrative''} and valued that the web reviews cited their sources. \emph{P5} liked that the chatbot \textit{``didn't try telling me what a wonderful choice I made.''} Trust was not unconditional, however. \emph{P4} would \textit{``check the information it was giving me against another source''} the first few times and noted that no AI is without \textit{``blind spots.''} Both \emph{P4} and \emph{P5} also raised the possibility of being steered toward more expensive options in realistic commercial contexts.

\section{Discussion and Design Implications}\label{sec:discussion}
\textbf{From transparent to usable.} Our findings suggest that transparency by design was necessary but not sufficient for understanding. The advisor surfaced its ranking rationale, labeled its AI-generated content, and reflected back inferred requirements, yet its ``Why this ranking?'' explanation was the single most severe problem. The confusion traced back to the explanation's most literal element: the loss (penalty) values were presented as penalties rather than as matches, whereas the positively framed radar chart fared better. Exposing exact system-internal values did not, in itself, produce understanding, echoing McKay's~\cite{mckay2022transparency} point that a fully explainable system can still leave people unable to make sense of a decision, and Tintarev and Masthoff's~\cite{tintarev2012explanations} point that transparency supports trust only when the exposed workings give users confidence. The quantitative results are consistent: ease was lowest for the ranking-explanation task, while satisfaction remained high.

\textbf{Control and a legible process.} Some participants also wanted more control over the advisor's inputs (filters, sorting) and outputs (explanation detail, category-specific attributes) than the conversational interface gave. A conversational advisor takes on the bulk of the search, so using it means ceding some direct control, re-enacting the direct-manipulation-versus-agents tension~\cite{shneiderman1997dm}. Human-Centered AI answers this by combining high human control with high automation~\cite{shneiderman2022hcai}, the mixed-initiative goal of coupling automation with direct manipulation~\cite{horvitz1999mixed}. Our advisor provides such automation and uses dialogue to resolve uncertainty about preferences, but lacks the direct-manipulation half, which our design implications supply. It also reflects back how it interpreted each request to keep the process legible and support self-efficacy~\cite{shneiderman2022hcai, Lee2026}, yet the fact that the ranking could still penalize a requirement the user never stated shows this feedback did not go far enough.

\textbf{Effort-driven, conditional adoption.} The general willingness to adopt held despite the problems above. Even \emph{P4}, the most critical of the explanation, would \textit{``give it a go.''} \emph{P5}, who reported the sample's highest self-rated laptop knowledge (6/7), saw less need for it, noting \textit{``I rarely ever buy something without already knowing exactly what I want.''} For the rest, the appeal was the effort saved by the advisor's scaffolding of the search~\cite{xu2023chatgpt, potiagalova2025conversational}. \emph{P3}, for example, found the advisor's clarifying questions \textit{``intuitive in terms of trying to understand the request.''} This is consistent with amplifying users' abilities while they retain the decision~\cite{shneiderman2022hcai}. But it was conditional on trust: strengthened by the non-pushy tone \emph{P5} valued (a contrast to human-feedback-tuned sycophancy~\cite{sharma2024sycophancy}) and weakened by the opaque penalties and low-credibility reviews. Our advisor performs only short, predefined operations, but we expect this delegation-versus-control tension to grow as autonomous agents take on longer, higher-stakes work with less oversight~\cite{Shah2025}. That participants wanted more control even in our context aligns with evidence that active use rather than passive reliance preserves users' sense of agency~\cite{Lee2026}, which our direct-manipulation implications address.

\textbf{Design implications.} We propose four changes to test. (i)~For the ranking explanation, carry the radar's positive, match-based framing over to the penalty table so that the table presents how well each option matches instead of how much it is penalized. The overlaid radar should give way to a representation that stays legible for three items. Inferred requirements should be surfaced for users to confirm or correct, and the ordering should signal more clearly that the leftmost recommendation is the best match. (ii)~For control, pair conversational elicitation with lightweight direct-manipulation controls, such as sliders whose adjustment updates the results immediately~\cite{shneiderman2022hcai}, alongside user-directed sorting and an adjustable level of detail in the ranking explanation. (iii)~For discoverability, clarify what selecting a card does and signal when new content has appeared below the current view. (iv)~For trust, keep the provenance labels and source citations, and let users steer which kinds of sources the web search draws on, for example favoring expert review sites over the crowd-sourced opinions that \emph{P4} distrusted.

\section{Conclusion, Limitations, and Future Work}\label{sec:conclusion}
We studied a transparent conversational product advisor, contributing severity-rated usability problems and design implications. The main lesson is that introducing transparency did not ensure it worked here: the central transparency feature was also the source of the most severe problem.

This study has several limitations. It was formative, with a small, all-UK sample, so the results are design insight, not frequency estimates. We evaluated a single system with no comparison condition, with possible novelty effects. Because laptops are defined largely by structured specifications, the ranking-explanation results may generalize more readily to similarly technical products than to experiential ones. The moderated setting meant some features were reached only after a moderator prompt, so use reflects guided rather than spontaneous discovery, though the need for prompting itself evidences the discoverability problem. Our measures were subjective, with no task-success or behavioral-trust metrics, so the design conclusions rest on self-report and observation. Single-analyst coding risks an evaluator effect~\cite{jacobsen1998evaluator} that a second coder would mitigate~\cite{mcdonald2019reliability}.

We see three directions for future work. First, the ranker covers only structured specification fields. Adding semantic retrieval over descriptions and reviews, as a bounded re-ranking term after the hard constraints, would let experiential attributes inform the ranking and its explanation while keeping the ranking deterministic. The explanation could then report semantic match scores and evidence quotes, extending the approach to less structured products. Second, the design implications point toward supplying the missing direct-manipulation half through filters, sliders, and user-directed sorting. Finally, a summative or comparative evaluation would test whether these changes improve comprehension and trust, as well as the hypothesis our results suggest: that such advisors are most helpful for users with low-to-medium domain knowledge and vague needs.

\begin{acks}
During the preparation of this work, the authors used a generative AI assistant (a large language model) to help structure the manuscript, draft and edit text, organize and cross-check the related work, prepare the tables and figures, and support the thematic coding of the study data. All resulting content was reviewed, verified, and where necessary corrected by the authors, who take full responsibility for the content of this paper.

This work was supported by the German Research Foundation (DFG) as part of the ``VACOS 2'' project (no. 388815326). We thank our study participants and our anonymous reviewers for their helpful feedback. We would also like to thank Jan Lattenkamp for his assistance with the technical implementation.
\end{acks}

\bibliographystyle{ACM-Reference-Format}
\bibliography{bibliography}

\end{document}